\documentclass[a4paper,notitlepage]{revtex4-1}
\usepackage{graphicx}
\usepackage{color}
\usepackage{amsmath,bm}
\usepackage{amssymb}
\usepackage{hyperref}

\usepackage[per-mode=symbol]{siunitx}   
\usepackage[version=3]{mhchem}

\DeclareSIUnit\intensity{\watt\per\centi\meter\squared}
\DeclareSIUnit\fieldstrength{\volt\per\centi\meter}
\DeclareSIUnit\kfieldstrength{k\volt\per\centi\meter}

\newcommand{\costT}{\ensuremath{\langle\langle\cos\theta\rangle\rangle_{T}}}%

\newcommand{\cost}{\ensuremath{\langle\cos\theta\rangle}}%
\newcommand{\degree}{\ensuremath{^\circ}}%
\newcommand{\eg}{e.\,g.}%
\newcommand{\Estatabs}{\ensuremath{\text{E}_{\textup{s}}}}%
\newcommand{\ie}{i.\,e.}%
\newcommand{\Ialign}{\textup{I}_{0}}%
\newcommand{\Nuptotal}{\ensuremath{\text{N}_\text{up}/\text{N}_\text{tot}}}%
\newcommand{\ffstate}[3]{\ensuremath{\left|#1,#2,#3\right>}}%
\newcommand{\pstate}[3]{\ensuremath{\left|#1,#2,#3\right>^\beta}}%
\newcommand{\pstatetreinta}[3]{\ensuremath{\left|#1,#2,#3\right>^{30}}}%
\newcommand{\ppstatetreinta}[3]{\ensuremath{\left|#1,#2,#3\right>_\text{p}^{30}}}%
\newcommand{\rpstate}[3]{\ensuremath{{}^\beta\left<#1,#2,#3\right|}}%
\newcommand{\ppstate}[3]{\ensuremath{\left|#1,#2,#3\right>_\text{p}}^\beta}%
\newcommand{\rppstate}[3]{\ensuremath{{}^\beta{}_\text{p}\left<#1,#2,#3\right|}}%
\newcommand{\pstateparallel}[3]{\ensuremath{\left|#1,#2,#3\right>^0}}%
\newcommand{\ppstateparallel}[3]{\ensuremath{\left|#1,#2,#3\right>_\text{p}^0}}%

\makeatletter
\def\subsubsection{\@startsection{subsubsection}{3}{10pt}{-1.25ex plus -1ex minus -.1ex}{0ex plus 0ex}{\normalsize\bf}}
\def\paragraph{\@startsection{paragraph}{4}{10pt}{-1.25ex plus -1ex minus -.1ex}{0ex plus 0ex}{\normalsize\textit}}
\renewcommand\@biblabel[1]{#1}
\renewcommand\@makefntext[1]%
{\noindent\makebox[0pt][r]{\@thefnmark\,}#1}
\DeclareRobustCommand\onlinecite{\@onlinecite}
\def\@onlinecite#1{\begingroup\let\@cite\NAT@citenum\citealp{#1}\endgroup}
\def\tagform@#1{\maketag@@@{\ignorespaces#1\unskip\@@italiccorr}}
\let\orgtheequation\theequation
\def\theequation{(\orgtheequation)}
\makeatother

\begin{document}
\title{Mixed-field orientation of a thermal ensemble of linear polar molecules}

\author{Juan J. Omiste and Rosario Gonz\'alez-F\'erez}
\address{{Instituto Carlos I de F\'{\i}sica Te\'orica y Computacional and
      Departamento de F\'{\i}sica At\'omica, Molecular y Nuclear, Universidad de Granada, 18071
      Granada, Spain}}

\email{rogonzal@ugr.es}

\begin{abstract}
We present a theoretical study of the impact of an electrostatic field combined with nonresonant linearly
polarized laser pulses on the rotational dynamics of a thermal ensemble of linear molecules.
We solve the time-dependent Schr\"odinger equation within the rigid rotor approximation for several
rotational states. Using the carbonyl sulfide  (OCS)
molecule as a prototype, the mixed-field orientation of a thermal
sample is analyzed in detail for experimentally accessible static field
strengths and laser pulses.
We demonstrate that for the characteristic field configuration used in current mixed-field
orientation experiments, a significant 
orientation is obtained for rotational temperatures below $0.7$~K or using stronger dc fields. 
\end{abstract}

\maketitle

\section{Introduction}
The mixed-field orientation of polar molecules via the interaction
with an electric field and a nonresonant laser field is a widespread technique to produce 
samples of oriented molecules. 
This method was  proposed by 
Friedrich and Herschbach~\cite{friedrich:jcp111,friedrich:jpca103},
and is based on the dc-field induced coupling between the 
nearly degenerate pair of states with opposite parity forming the tunneling
doublets in the strong laser field regime.
A recent experimental and theoretical study
has proven that under ns laser pulses the weak dc field
orientation is not, in general, adiabatic, 
and that a time-dependent description of the mixed-field orientation process is required
to explain the experimental results~\cite{nielsen:prl2012,omiste_nonadiabatic_2012}. 
Thus, depending on the field configuration,  the orientation of a
rotational state could  be significantly smaller than the adiabatic prediction. 
In addition, not all the states present a right-way orientation, and some of them are
antioriented.

In a thermal ensemble of molecules,
the combination of these right- and wrong-way oriented states  
gives rise to a weakly oriented molecular beam~\cite{sakai:prl_90,Buck:IRPC25:583}. 
An enhancement of the orientation could be achieved by employing either 
lower rotational temperatures or 
quantum-state selected molecular beams. 
By using inhomogeneous electric fields, the amount of populated states
is significantly reduced creating a quantum-state selected
molecular beam, and achieving with this beam an unprecedented degree of 
orientation~\cite{ghafur_impulsive_2009,kupper:prl102,kupper:jcp131}.
Cold molecular beams, with typical temperatures of the
order of $1$~K, are created in supersonic expansions
of molecules seeded in an inert atomic carrier gas~\cite{even:8068}.
Depending on the rotational constant,  the molecules could still  be 
distributed over a large number of rotational states in these thermal ensembles. 
In the present work, we investigate the mixed-field orientation of a thermal sample of polar molecules as the
rotational temperature is varied. Our aim is to find the temperature at which 
the thermal ensemble shows a similar orientation as the quantum-state selected molecular beam.
 
Herein, we consider a  polar linear molecule exposed to an electric field combined
with a nonresonant laser pulse, and provide a detailed theoretical analysis of the mixed-field
orientation of a thermal sample of this molecule.
To do so,  we solve the time-dependent Schr\"odinger 
equation within the rigid rotor approximation for a large set of rotational states. 
Taking as prototype example
the OCS molecule, we explore the mixed-field orientation as a function of the rotational 
temperature of the thermal sample for several experimental field configurations.
We show that to achieve a significant orientation,
rotational temperatures around  $0.6$~K and $1$~K are required if either 
a weak or strong dc fields are applied, respectively. 
We also present the orientation of individual states and, for some of them, 
analyze  the projections of the time-dependent wave functions  on
the corresponding adiabatic basis.  

The paper is organized as follows: In \autoref{sec:hamiltonian} we describe
the Hamiltonian of the system and the orientation of a molecular thermal ensemble. 
The mixed-field orientation of the thermal ensemble 
as a function of the rotational temperature 
is  analyzed in \autoref{sec:results}. 
The conclusions are given in \autoref{sec:conclusions}.

\section{The Hamiltonian and the orientation of a thermal ensemble}
\label{sec:hamiltonian}

We consider a polar linear molecule  exposed to a homogeneous
static electric field and a nonresonant linearly polarized laser pulse. 
In the framework of the rigid rotor approximation,
the Hamiltonian of this system reads  
\begin{equation}
  \label{eq:hamiltonian}
  H(t)=H_r+H_{\textup{s}}(t)+H_{\textup{L}}(t),
\end{equation}
where $H_r$ is the field-free Hamiltonian
\begin{equation}
  \label{eq:hr}
  H_r=B\mathbf{J}^2,
\end{equation}
with $\mathbf{J}$ being  the total angular momentum operator and $B$ the
rotational constant. The
interactions with the electric and laser fields are
$H_{\textup{s}}(t)$ and $H_{\textup{L}}(t)$,  respectively.

The dc field $\mathbf{E}_{\textup{s}}(t)$ forms an angle $\beta$ with the $Z$-axis and is contained in the $XZ$-plane
of the laboratory fixed frame (LFF) $(X,Y,Z)$.
The dipole coupling with this field  reads 
\begin{equation}
  \label{eq:hs}
  H_{\textup{s}}(t)=-\boldsymbol{\mu}\cdot\mathbf{E}_{\textup{s}}(t) 
=-\mu \Estatabs(t)\cos\theta_{\textup{s}}
\end{equation}
with  $\mathbf{E}_{\textup{s}}(t)=\Estatabs(t)(\sin\beta \hat{X}+\cos\beta \hat{Z})$, 
and $\Estatabs(t)$ being the electric  field strength.
The angle between the dipole moment $\boldsymbol{\mu}$ and $\mathbf{E}_{\textup{s}}(t)$ is 
$\theta_{\textup{s}}$, 
and 
$\cos\theta_{\textup{s}}=\cos\beta\cos\theta+\sin\beta\sin\theta\cos\phi$.
The angles $\Omega=(\theta,\phi)$ are the Euler angles, which relate 
the  laboratory and molecular fixed frames. 
The molecule fixed frame (MFF) $(X_M,Y_M,Z_M)$  is defined so that the molecular
permanent dipole moment $\boldsymbol{\mu}$  is parallel to the $Z_M$-axis.
Based on the mixed-field orientation experiments~\cite{kupper:prl102,kupper:jcp131,nielsen:prl2012},
the dc field is switched on first increasing its 
strength linearly with time. We ensure that this turning-on process is adiabatic, and  once the 
maximum strength $\Estatabs$ is achieved, it is kept constant.

The polarization of the nonresonant laser field is taken parallel to the $Z$-axis.  Thus, 
the interaction of the nonresonant laser field with the 
molecule can be written as~\cite{seideman2006}
\begin{equation}
  \label{eq:hl}
  H_\textup{L}(t)=-\cfrac{\textup{I}(t)}{2c\epsilon_0}\Delta\alpha\cos^2\theta,
\end{equation}
where $\Delta\alpha$ is the polarizability anisotropy, $\textup{I}(t)$ is the intensity of the
laser, $c$ is the speed of light and $\epsilon_0$ is the dielectric
constant. 
Note that in~\autoref{eq:hl}
the term $-\alpha_\perp \textup{I}(t)/2c\epsilon_0$ has been neglected because it represents only a shift in the energy. 
The laser  is a Gaussian  pulse with intensity 
$\textup{I}(t)=\Ialign\exp\left(-t^2/2\sigma^2\right)$, 
$\Ialign$ is the peak intensity, and  $ \sigma$ is related with the 
full width half maximum (FWHM) $\tau=2\sqrt{2\ln 2} \sigma $.  
When the nonresonant laser field is turned on   
the interaction due to this field is much weaker 
than the coupling with  the dc field.

The time-dependent Schr\"odinger equation 
associated to the Hamiltonian \eqref{eq:hamiltonian}
is solved by means of 
a second-order split-operator technique~\cite{feit:jcp82}, combined with 
the discrete-variable and finite-basis representation
methods for the angular coordinates~\cite{bacic:arpc89,corey:jcp92,offer:10416,sanchezmoreno:PRA.2007}. 
The  basis is formed by the spherical harmonics $Y_{JM}(\Omega)$, which are the 
eigenstates of the field-free Hamiltonian \eqref{eq:hr}. 
$J$ and $M$ are the rotational and magnetic quantum numbers, respectively. 
At time $t$, the  time-dependent states will be labelled as 
\pstate{J}{M}{l}$_t$
with $l=e$ and $o$ indicating even or odd parity
with respect to the $XZ$-plane, respectively.
The labels $J$, $M$ and $l$ refer to the field-free quantum numbers to which they are adiabatically
connected and they depend on the way the fields are turned on~\cite{hartelt_jcp128}.

We consider a thermal sample of molecules  and  
investigate its mixed-field orientation  at $t=0$ once the peak intensity 
$\Ialign$ has been achieved. 
For a  rotational temperature $T$, the orientation of a thermal distribution is given by
\begin{equation*}
\label{eq:orientation_thermal}
\langle\cost\rangle_{T}=
\sum_{J=0}^\infty\sum_{M=-J}^{J} W_{J}^T\cost_{JM}
\end{equation*}
where the orientation of the field-dressed state \pstate{J}{M}{l}$_0$
is $\cost_{JMl}={}_0$\rpstate{J}{M}{l}$ \cos\theta$\pstate{J}{M}{l}$_0$.
The thermal weight of the field-free state \ffstate{J}{M}{l} is 
\begin{equation}
\label{eq:weight_thermal}
W_{J}^T=\frac{e^\frac{-J(J+1)B}{k_BT}}{W^T} \qquad  W^T=\sum_{J=0}^\infty(2J+1)e^\frac{-J(J+1)B}{k_BT}
\end{equation}
with $k_B$ being the Boltzman constant.

In many mixed-field orientation experiments,  the degree of
orientation is measured  by the ion imaging method \cite{kupper:prl102,kupper:jcp131}.
The up/down symmetry of the 2D-images of the ionic fragments 
is experimentally quantified by the ratio 
$\Nuptotal$, with $\text{N}_\text{up}$ being the amount of ions in the upper part
of the screen plane, and $\text{N}_\text{tot}$ the total number of detected ions.
In order to compare with the experimental results~\cite{nielsen:prl2012}, 
we also compute the orientation ratio
$\Nuptotal$, of this thermal sample
  on a  2D screen perpendicular to the electric field axis. 
This is defined  as 
\begin{equation*}
\frac{\text{N}_{\textup{up}}}{\text{N}_{\textup{tot}}}=\sum_{J}\sum_{M=-J}^{J} W_{J}^T
\frac{\text{N}_{\textup{up}}^{JM}}{\text{N}_{\textup{tot}}^{JM}}
\end{equation*}
where
\begin{equation}
\label{eq:Nups}
\text{N}_{\textup{up}}^{JM}=\int_{y^2+z^2\le 1}\int_{z\ge0} P_{JM}(y,z) \,dydz,
\end{equation}
and
\begin{equation*}
\label{eq:Ntotal}
\text{N}_{\textup{tot}}^{JM}=\int_{y^2+z^2\le 1} P_{JM}(y,z) \,dydz 
\end{equation*}
with $P_{JM}(y,z)$ being the  projection  on a 2D screen perpendicular to the electric field axis 
 of the probability density  associated to the state \pstate{J}{M}{l}$_0$~\cite{omiste:pccp2011},
which includes the alignment selectivity of the probe laser.
$y$ and $z$ are the  abscissa and ordinate of a 2D coordinate system centered
 on the screen, 
 due to their relation with the Euler angles $(\theta,\phi)$ their values are restricted to 
 $y^2+z^2\le 1$~\cite{omiste:pccp2011}.

To rationalize the mixed-field orientation results and illustrate the adiabaticity of this process,
the time-dependent  wave function is projected on the  
field-dressed adiabatic states
\begin{equation}
 \label{eq:projection_psit}
\pstate{J}{M}{l}_t=\sum_{j=0}^N\sum_{m_j=-j}^{j}C_{j m_jl'}(t)\ppstate{j}{m_j}{l'}
\end{equation}
with $  C_{jm_jl'}(t)=\rppstate{j}{m_j}{l'} J Ml\rangle^\beta_t$.
This adiabatic basis is formed by the eigenstates $\ppstate{j}{m_j}{l}$
of the adiabatic Hamiltonian, \ie, the Hamiltonian \ref{eq:hamiltonian} with
constant electrostatic field $\Estatabs$
and constant laser intensity $\textup{I}=\textup{I}(t)$.
For each time $t$, the time-independent Schr\"odinger equation 
is solved by expanding the wave function in a basis formed by linear combinations of spherical harmonics that respects
the symmetries of the system.
Note that for \pstate{J}{M}{l}$_0$, the closer  $ | C_{JM l}|^2$ to one the more adiabatic is the
mixed-field orientation process.

\section{Results}
\label{sec:results}
\begin{figure}[t]
\begin{minipage}[h]{30pc}
\begin{minipage}[h]{14pc}
\includegraphics[angle=0,width=14pc]{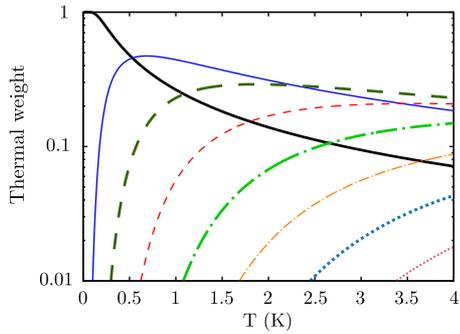}
\end{minipage}\hspace{2pc}%
\begin{minipage}[h]{14pc}
\caption{For OCS, thermal weights as a function of the temperature for several $J$-manifolds:  
$J=0$ (thick solid line),
$J=1$ (thin solid line),
$J=2$ (thick dashed line),
$J=3$ (thin dashed line),
$J=4$ (thick dot-dashed line),
$J=5$ (thin dot-dashed line),
$J=6$ (thick dotted line)
and $J=7$ (thin dotted line).
  \label{fig:weights_jm}}
\end{minipage}
\end{minipage}
\end{figure}

In this work, we use the OCS molecule as prototype.  
The rotational constant of OCS is $B=0.20286$ cm$^{-1}$, 
the permanent dipole moment $\mu=0.71$~D and the polarizability anisotropy
$\Delta\alpha=4.04$ \AA$^3$. 
In \autoref{fig:weights_jm}, we present the thermal weights 
of several rotational manifolds  $(2J+1)W_{J}$, see  \autoref{eq:weight_thermal}. 
Due to the large rotational constant of OCS, the field-free energy splittings are large, and then, the
thermal samples with $T\lesssim 1$~K are dominated by the $J=0$ and $J=1$ manifolds.
Indeed, the relative weights of the states with $J=0$ and $J=1$ are $W_0=47.8\%$ and $W_1=44.7\%$ at $T=0.5$~K,
and $W_0=99.1\%$ and $W_1=0.9\%$  at $T=0.1$~K.  
In our calculations, the thermal sample includes rotational states with $J\le 9$, and we have ensured that
the contribution of higher excitations can be neglected.

\begin{figure}[t]
\centering
 \includegraphics[width=36pc]{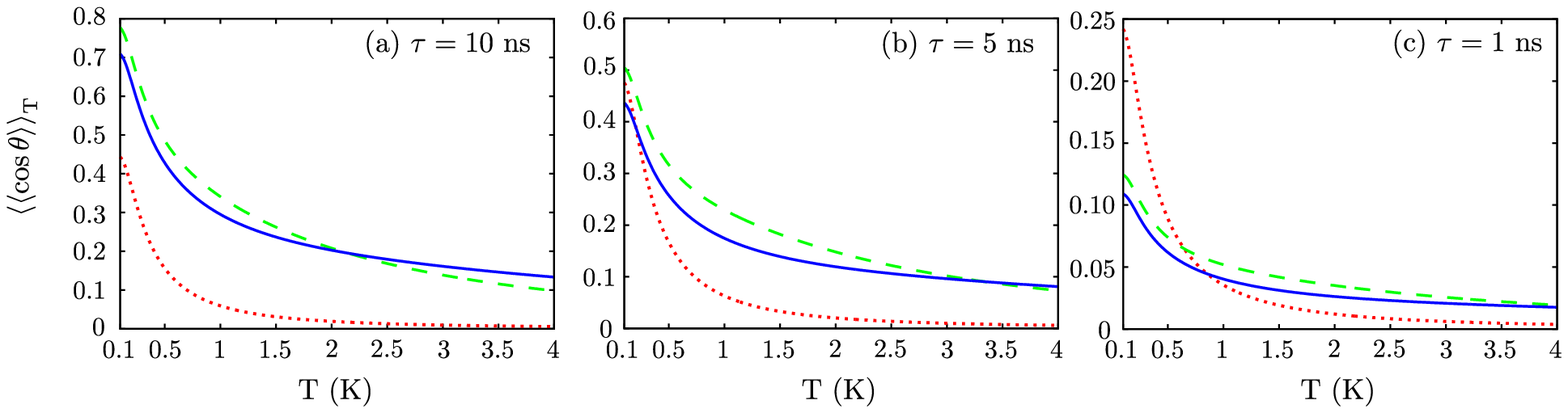}
\caption{\label{fig:fc_300_10_0_I} 
Orientation of a OCS thermal sample  $\costT$ as a function of the temperature for Gaussian 
pulses with $\tau=10$~ns, $\tau=5$~ns, and $\tau=1$~ns and peak intensities
$\Ialign=\SI{e12}{\intensity}$  (thick solid line), $\Ialign=\SI{5e11}{\intensity}$ (dashed line)
and 
$\Ialign=\SI{e11}{\intensity}$
(dotted line). The field  configuration is  $\Estatabs=\SI{300}{\fieldstrength}$ and $\beta=0\degree$. }
\end{figure}

We first consider the OCS molecules exposed to an electric field and linearly polarized laser pulse, 
with both fields  parallel to the LFF $Z$-axis.
For several field configurations, 
we present in \autoref{fig:fc_300_10_0_I} the orientation cosine of the thermal 
ensemble as a function of the temperature for 
$\Estatabs=\SI{300}{\fieldstrength}$. Note the 
different scales used in each panel. 

\begin{figure}[h]
\centering
\includegraphics[width=30pc]{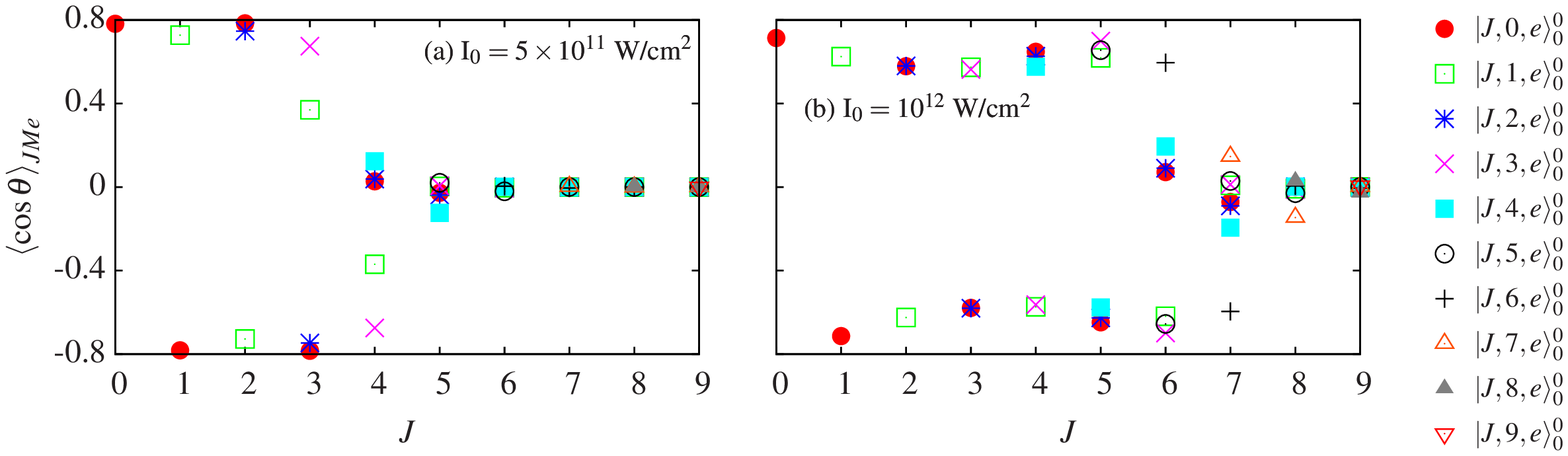}
\caption{\label{fig:cos_JM_300_10_0} 
Orientation cosines $\cost_{JMl}$ at $t=0$ of the states 
\pstateparallel{J}{|M|}{e}$_0$ 
versus the field-free rotational quantum number $J$. 
The Gaussian pulses have  $\tau=10$~ns,  and peak intensities
(a) $\Ialign=\SI{5e11}{\intensity}$ and
(b) $\Ialign=\SI{e12}{\intensity}$.
 The field  configuration is  $\Estatabs=\SI{300}{\fieldstrength}$ and $\beta=0\degree$. 
}
\end{figure}

For this weak dc field, a significant orientation 
is only achieved  if the rotational temperature  is below $0.5$~K,
and the Gaussian pulse has $\tau=10$~ns, \eg, for the peak intensities  
$\Ialign=\SI{e12}{\intensity}$ and $\SI{5e11}{\intensity}$  we obtain  $\costT\gtrsim0.5$. 
Using $1$~ns Gaussian pulse, the orientation of the thermal sample is very small because 
the rotational states are weakly oriented, for instance, they satisfy  $|\cost_{JMl}|<0.13$ for 
$\Ialign=\SI{e12}{\intensity}$ and $\Ialign=\SI{5e11}{\intensity}$; whereas
for $\Ialign=\SI{e11}{\intensity}$,
we obtain $\cost_{00e}=0.24$ for the ground state. 
For these three FWHM,  we encounter that 
a pulse with peak intensity $\Ialign=\SI{5e11}{\intensity}$ gives rise to a larger orientation
than one with $\Ialign=\SI{e12}{\intensity}$, this is counterintuitive
to what is expected in the adiabatic limit. 
This phenomenon can be explain by 
the non-adiabaticity of the mixed-field orientation  
process~\cite{nielsen:prl2012,omiste_nonadiabatic_2012}, and can be rationalized in terms of the  
orientation of the individual levels.
In \autoref{fig:cos_JM_300_10_0}, we present the orientation cosine of the field-dressed states
\pstateparallel{J}{|M|}{e}$_0$ at $t=0$  for two $10$~ns Gaussian pulses 
with $\Ialign=\SI{e12}{\intensity}$ and $\Ialign=\SI{5e11}{\intensity}$. 
In these plots, we observe that 
the levels \pstateparallel{J}{M}{e}$_0$--\pstateparallel{J+1}{M}{e}$_0$,
which form a pendular doublet, are oriented and antioriented, respectively. 
The $\SI{5e11}{\intensity}$ pulse is not strong enough to affect the rotational dynamics in the 
excited rotational states with $J\ge5$.
The pulse with the strongest intensity $\Ialign=\SI{e12}{\intensity}$
provokes a large orientation on highly excited states with $J\le 7$. 
However, for the levels with $J\le 3$, \ie,  those that are important on the cold regime, 
the $\SI{5e11}{\intensity}$ pulse gives rise to a larger orientation
compared to the $\SI{e12}{\intensity}$ one.
In the parallel field configuration, the population transfer between the two levels forming 
the doublets in the pendular regime is the only source of
nonadiabatic effects in the field-dressed 
dynamics~\cite{nielsen:prl2012,omiste_nonadiabatic_2012}.
For these levels,  the population transfer to the neighboring state 
as the pendular pair is formed is the largest for the strongest laser.
For the ground state, at $t=0$  we obtain 
that the population of the adiabatic state $\ppstateparallel{0}{0}{e}$ is 
$|C_{00e}|^2=0.87 $ and $0.91$
with $\Ialign=\SI{e12}{\intensity}$
and  $\Ialign=\SI{5e11}{\intensity}$, respectively.
As a
consequence, the orientation is smallest for $\Ialign=\SI{e12}{\intensity}$, 
and, therefore, the thermal ensemble is less oriented. 
By increasing the temperature, the contribution of excited rotational states becomes 
important, and the thermal ensemble in a  $\Ialign=\SI{e12}{\intensity}$ pulse shows the largest orientation.

\begin{figure}[h]
\centering
\includegraphics[width=36pc]{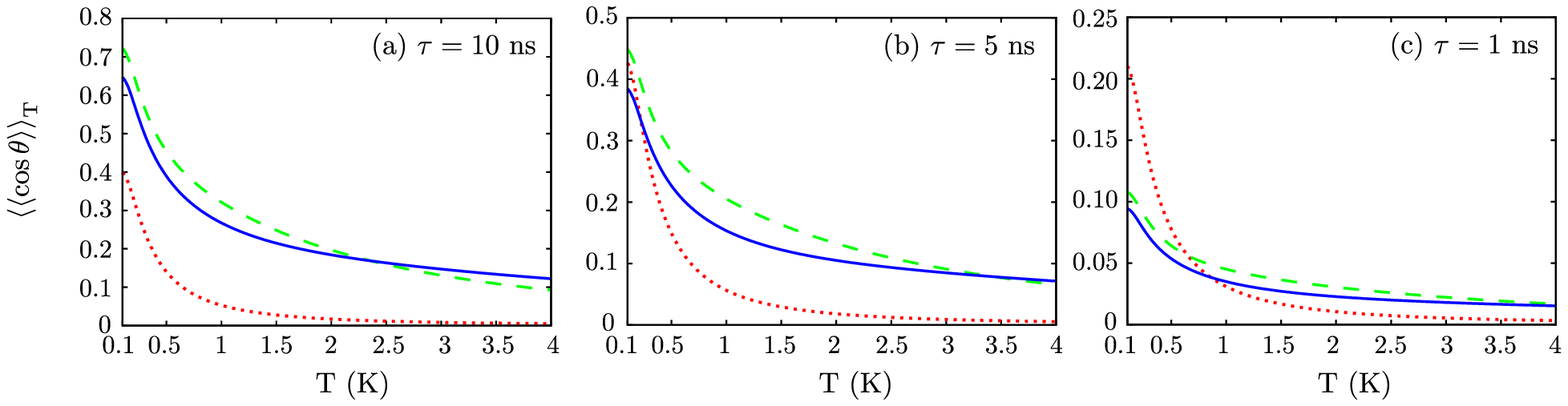}
\caption{\label{fig:fc_300_10_30_I}
Same as \autoref{fig:fc_300_10_0_I} but for $\beta=30\degree$. }
\end{figure}

\begin{figure}[h]
\centering
\includegraphics[width=30pc]{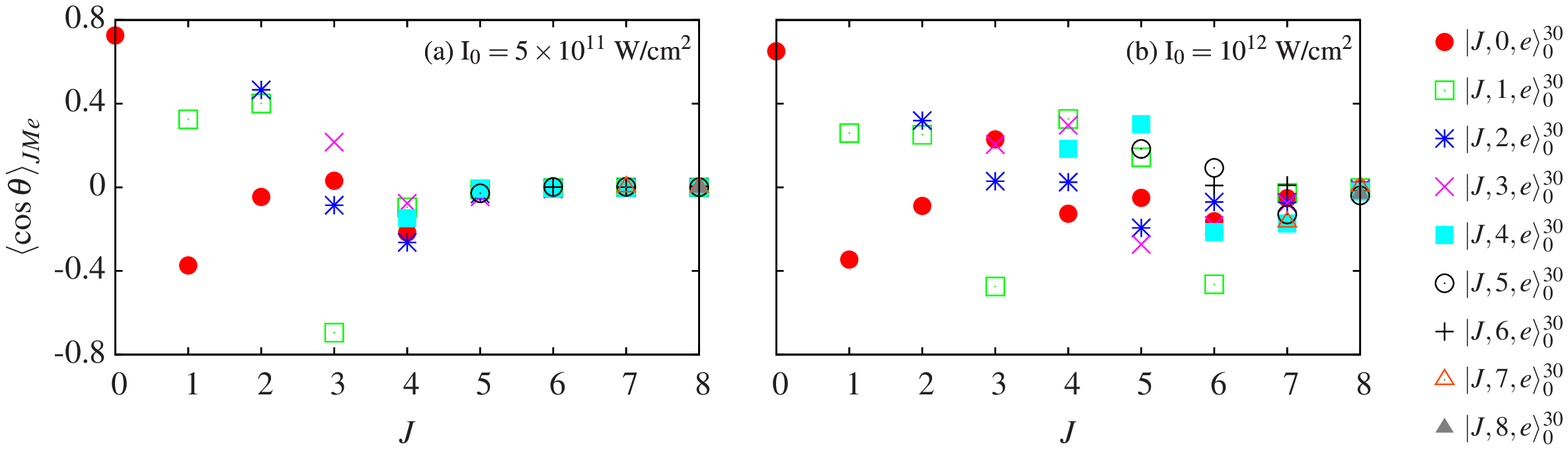}
\caption{\label{fig:cos_JM_300_10_30} 
For the states 
\pstatetreinta{J}{|M|}{e}$_0$, 
orientation cosines $\cost_{JMl}$ at $t=0$ versus the field-free rotational quantum number
$J$. 
The Gaussian pulses have $\tau=10$~ns  and  peak intensities
(a) $\Ialign=\SI{5e11}{\intensity}$ and
(b) $\Ialign=\SI{e12}{\intensity}$.
 The field  configuration is  $\Estatabs=\SI{300}{\fieldstrength}$ and $\beta=30\degree$. }
\end{figure}

Now, we consider that the electric field is tilted an angle $\beta=30\degree$ with respect to the
polarization axis of the laser pulse, that is the LFF $Z$-axis.
For several field configurations, we present in \autoref{fig:fc_300_10_30_I} 
the orientation cosine of the thermal  ensemble as a function of the temperature for 
$\Estatabs=\SI{300}{\fieldstrength}$. 
Compared to the parallel field case, the orientation is reduced.
For tilted fields, there are two main sources of nonadiabatic effects
in the  field-dressed dynamics:
i) the transfer of population taking place when the quasidegenerate pendular doublets are formed
as the laser intensity is increased;
ii) at weak laser intensities,  there is also  population transfer due to the splitting of the states 
within a $J$-manifold now having the same symmetry.
In addition, avoided crossings might be encountered as $I(t)$ is enhanced.
The diabatic or adiabatic character of these avoided crossings depends on the field configuration  
and on the state. 
Hence, for a certain field configuration, 
the orientation of the individual states is smaller for $\beta=30\degree$ than for $\beta=0\degree$.
This reduction of the orientation is illustrated 
for the rotational states
\pstateparallel{J}{M}{e}$_0$ 
in \autoref{fig:cos_JM_300_10_30}  for two $10$~ns Gaussian pulses. 
For $\Ialign=\SI{5e11}{\intensity}$, only the states \pstateparallel{0}{0}{e}$_0$ and
\pstateparallel{3}{1}{e}$_0$ present a strong orientation with $|\cost_{JMl}|>0.6$, 
whereas for $\Ialign=\SI{e12}{\intensity}$ only the ground state is strongly oriented. 
The other levels present a moderate or even small orientation. 
Due to the population redistribution within a $J$-manifold at weak intensities, the 
two levels forming a pendular doublet do not possess the same orientation $|\cost_{JMl}|$ but in 
opposite directions as occurs in the parallel field configuration. 

\begin{figure}[t]
\centering
 \includegraphics[angle=0,width=30pc]{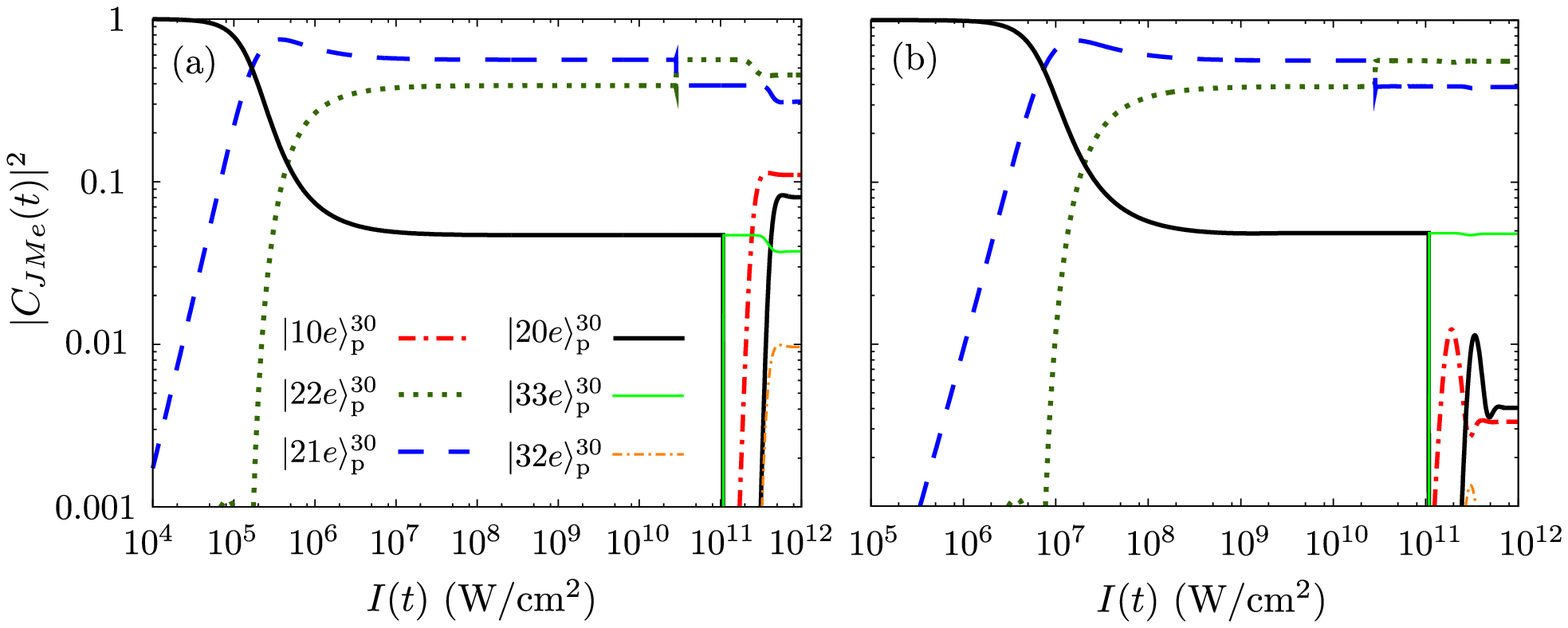}
\caption{\label{fig:coef_JM_300_2000} 
For the state 
\pstatetreinta{2}{0}{e}$_t$, we present the 
squares of the projections of the time dependent
wave function onto the adiabatic pendular states
versus  the laser intensity $\textup{I}(t)$, 
for dc field strengths
(a) $\Estatabs=\SI{300}{\fieldstrength}$ 
and 
(b) $\Estatabs=\SI{2}{k\fieldstrength}$.
The Gaussian pulse has $\tau=10$~ns and peak intensity
$\Ialign=\SI{e12}{\intensity}$, and the fields are tilted an angle $\beta=30\degree$.}
\end{figure}

For the state \pstatetreinta{2}{0}{e}$_t$, we illustrate its rotational dynamics 
by presenting the projections of the time-dependent wave function in terms of the adiabatic
states in~\autoref{fig:coef_JM_300_2000}(a) for a $10$~ns pulse with $\Ialign=\SI{e12}{\intensity}$.  
The switching on of the electric field has been adiabatic and the level
\ppstatetreinta{2}{0}{e} is the only one populated when the laser pulse is turned on. 
At weak laser intensities, the three states with the same symmetry in the $J=2$ manifold,
that is \ppstatetreinta{2}{0}{e}, \ppstatetreinta{2}{1}{e} and \ppstatetreinta{2}{2}{e}, 
are driven apart:  $|C_{20e}(t)|^2$ decreases as $\textup{I}(t)$ is increased,
whereas $|C_{21e}(t)|^2$  and $|C_{22e}(t)|^2$ increase.
For a wide range of laser intensities, these three coefficients keep their values constant.
Around $\textup{I}(t)\approx\SI{2.84e10}{\intensity}$, the states \ppstatetreinta{2}{1}{e} and
\ppstatetreinta{2}{2}{e} suffer an avoided crossings, which is crossed diabatically and the population
of these two adiabatic levels is interchanged. 
Another diabatic  avoided crossing is encountered around  $\textup{I}(t)\approx\SI{1.09e11}{\intensity}$,
and the involved states \ppstatetreinta{2}{0}{e} and \ppstatetreinta{3}{3}{e} interchanged their 
population. 
Upon further increasing $\textup{I}(t)$, the pendular doublets start to form, the coupling between the two
involved states increases, and there is a new population redistribution. In this figure, it is
appreciated how the different pendular doublets are formed sequentially according to their energy. 
The first one involves the states \ppstatetreinta{1}{0}{e} and \ppstatetreinta{2}{2}{e},
the next one \ppstatetreinta{2}{1}{e} and \ppstatetreinta{2}{0}{e}, and
the third one in this figure  \ppstatetreinta{3}{3}{e} and \ppstatetreinta{3}{2}{e}.
At $t=0$, the contribution of the adiabatic states to the field-dressed wave function is
$|C_{22e}(0)|^2 =0.11$, 
$|C_{22e}(0)|^2 =0.45$, 
$|C_{21e}(0)|^2 =0.31$, 
$|C_{20e}(0)|^2 =0.08$, 
$|C_{33e}(0)|^2 =0.04$ and 
$|C_{32e}(0)|^2 =0.01$.
As a consequence of this population redistribution, at $t=0$  
the state \pstatetreinta{2}{0}{e}$_0$ is weakly antioriented  $\cost_{20e}=-0.089$, 
whereas in the adiabatic prediction present a strong anti-orientation $\cost_{20e}=-0.886$.
Analogously,  other features of the
system such as the energy, alignment, and hybridization of
the angular motion are also affected by this population redistribution and do not resemble the
adiabatic results. 

\begin{figure}[t]
\centering
\includegraphics[width=36pc]{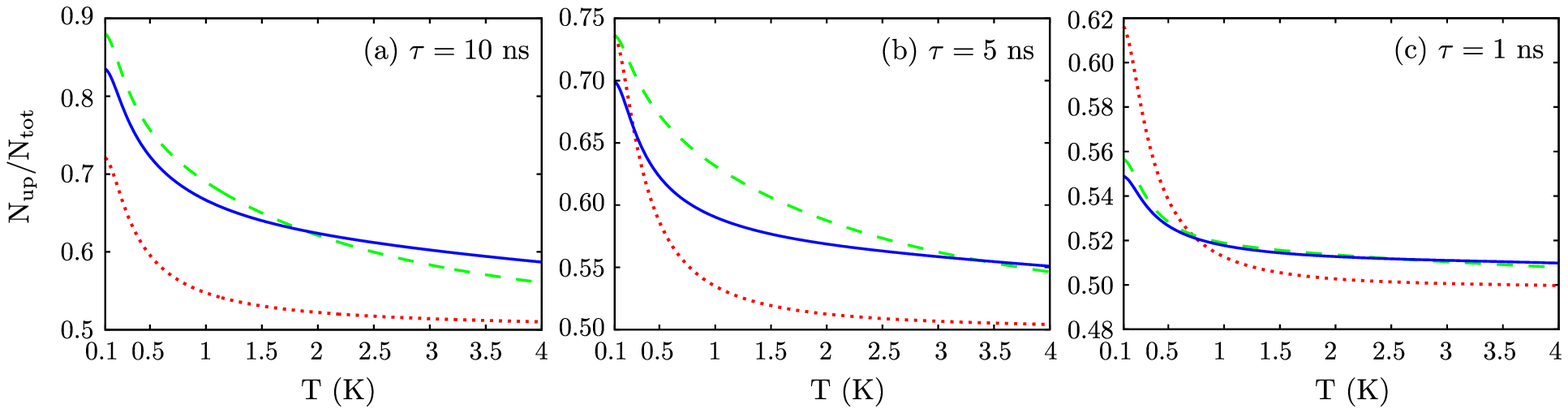}
\caption{\label{fig:FC_NUP_300_10_30_I} For the OCS thermal sample, we present the 
orientation ratio  $\Nuptotal$  as a function of the temperature for Gaussian pulses 
with FWHM $\tau=10$~ns, $\tau=5$~ns, and $\tau=1$~ns and peak intensities
$\Ialign=\SI{e12}{\intensity}$  (thick solid line), $\Ialign=\SI{5e11}{\intensity}$ (dashed line)
and $\Ialign=\SI{e11}{\intensity}$
(doted line). The field  configuration is  $\Estatabs=\SI{300}{\fieldstrength}$ and $\beta=30\degree$. }
\end{figure}

For $\beta=30\degree$, the orientation ratio $\Nuptotal$
is presented in \autoref{fig:FC_NUP_300_10_30_I}. 
To compute  $\Nuptotal$ we have used a probe laser 
linearly polarized along the vertical axis of the screen detector as in  
the experiments ~\cite{nielsen:prl2012}. 
In these results, we have neglected the volume effect~\cite{omiste:pccp2011},
we should mention that by including it the value of $\Nuptotal$ will be reduced.

In recent experiments~\cite{nielsen:prl2012}, 
for a state selected molecular beam of OCS, 
$92\%$ in \pstateparallel{0}{0}{e}$_0$,
$4\%$  in \pstateparallel{1}{1}{e}$_0$ and 
$4\%$  in \pstateparallel{1}{1}{o}$_0$, 
 an orientation ratio of $\Nuptotal=0.73$ was achieved 
using a $8$~ns YAG laser with $\Ialign=\SI{9.1e11}{\intensity}$,
$\Estatabs=\SI{286}{\fieldstrength}$ and $\beta=30\degree$.
Using a $10$~ns pulse, similar results for the orientation ratio of the thermal
ensemble are reached if the 
rotational temperature is  sufficiently low.
For instance, $\Nuptotal\gtrsim 0.73$ for $T\lesssim0.65$~K and $0.46$~K with 
peak intensities $\Ialign=\SI{5e11}{\intensity}$ and $\Ialign=\SI{e12}{\intensity}$, respectively.
At $T=0.65$~K, the field-free thermal ensemble is formed by
$38.56\%$ OCS in its ground state, $47.19\%$ in $J=1$ and $13\%$ in $J=2$; 
whereas for $T=0.46$~K, $51.08\%$ have $J=0$, 
$43.07\%$ $J=1$, and $5.7\%$ $J=2$. 
For $\tau=5$~ns, only when more than $95\%$ of OCS molecules are in the ground state 
and $\Ialign=\SI{5e11}{\intensity}$ we obtain a similar orientation ratio as in the experiment. 
By reducing the FWHM to $1$~ns, the orientation ratio is significantly reduced.

\begin{figure}[h]
\begin{minipage}[h]{30pc}
\begin{minipage}[h]{14pc}
\includegraphics[width=14pc]{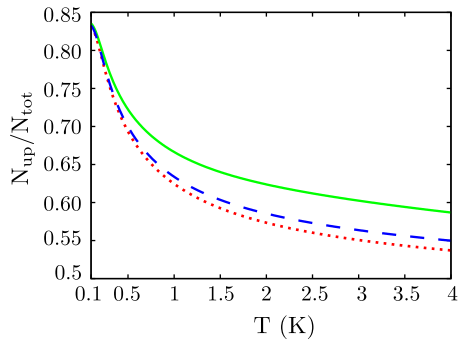}
\end{minipage}\hspace{2pc}%
\begin{minipage}[h]{14pc}
\caption{\label{fig:FC_NUP_300_10_30_1e12_probe} For a OCS thermal sample, 
we present the  orientation ratio  $\Nuptotal$ using a probe pulse 
linearly polarized along the vertical axis of the  screen  (thick solid line),
along the perpendicular axis to the screen (dashed line) and 
circularly polarized in a plane perpendicular to the screen  (doted line). 
The field parameters are  $\tau=10$~ns, $\Ialign=\SI{e12}{\intensity}$, 
$\Estatabs=\SI{300}{\fieldstrength}$ and $\beta=30\degree$.}
\end{minipage}
\end{minipage}
\end{figure}

An important ingredient to obtain realistic screen images and orientation ratios 
 is the alignment selectivity of
the probe laser, which depends on its  polarization~\cite{omiste:pccp2011}.
Here, we consider a thermal sample in a laser pulse with $\tau=10$~ns and 
$\Ialign=\SI{e12}{\intensity}$, and electric field 
$\Estatabs=\SI{300}{\fieldstrength}$ and $\beta=30\degree$.
In \autoref{fig:FC_NUP_300_10_30_1e12_probe}, we present 
its orientation ratio using the  probe pulse
with  three 
possible polarizations.
For a probe pulse linearly polarized parallel to the vertical axis of the screen, 
 $\Nuptotal$ is the largest  because such a pulse  favors the Coulomb explosion of the oriented 
 molecules.
 In contrast,  if  the probe pulse is linearly polarized perpendicular to screen,  the probability of the 
 Coulomb explosion for the oriented molecules is reduced, and, therefore, 
 $\Nuptotal$ presents the smallest values.
 The circularly polarized probe laser ensures that any molecule is ionized and detected
with the same probability independently of the angle $\beta$, and provides the intermediate values 
of $\Nuptotal$ for any temperature.
For a given state, 
there is no analytical relation between its orientation $\costT$ and the orientation ratio
$\Nuptotal $ of the 2D projection of its wave function, although the approximation 
$\Nuptotal\approx(1+\costT)/2$ could be used to obtain an estimation.
For instance,   a $ 0.29$~K thermal sample presents an orientation of
$\costT=0.506$, and  orientation ratios  
$\Nuptotal=0.757$ and $0.774$ for a probe laser linearly polarized perpendicular and parallel to 
the screen detector, respectively, and  $\Nuptotal=0.761$ for a circularly polarized one.
These results should be compared with the value $0.753$ given by this approximation, which provides
a lower bound for these three polarizations.

For parallel fields,  if the electric field strength is increased,  
the energy splitting in a pendular doublet is increased, and as a consequence, the degree of 
adiabaticity in the molecular mixed-field orientation is also enhanced.
However, this statement only holds for the ground state of the two irreducible representations 
if the fields are tilted.
For an excited rotational state,  a strong dc field does not ensure a  large orientation because the
coupling between levels with different field-free $M$ values becomes important, and this  affects 
the molecular dynamics.  
In contrast, for a weak dc field, the mixing between these states is so small that $M$ can be
considered as conserved. 

\begin{figure}[t]
\centering
\includegraphics[width=27pc]{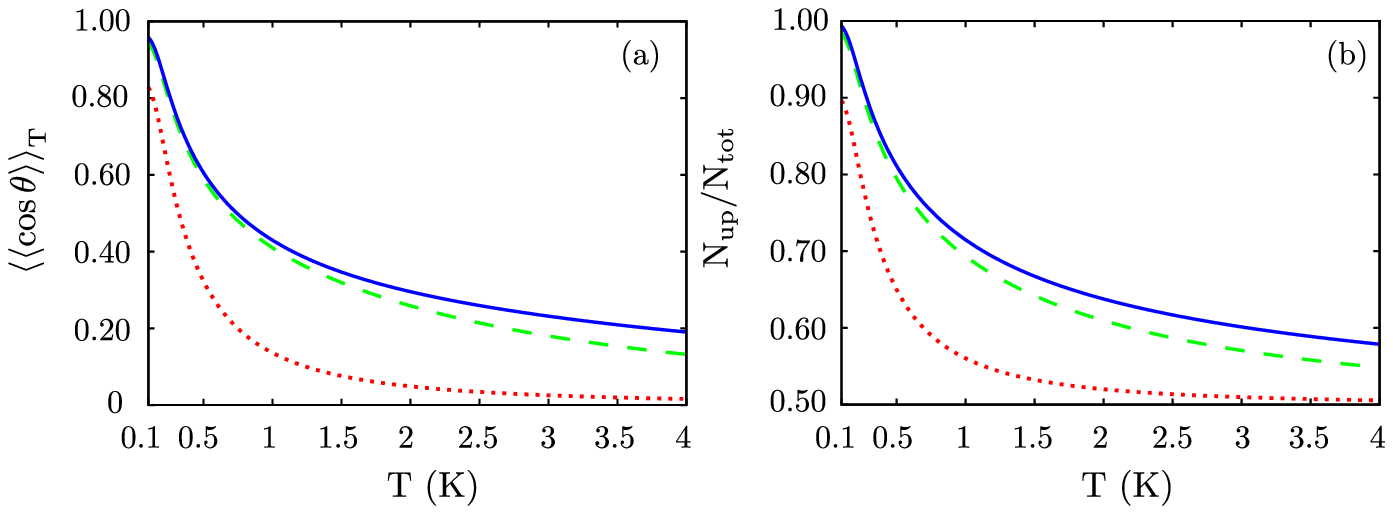}
\caption{\label{fig:fc_2000_10_30_I}
For a OCS thermal sample, we present 
(a) $\costT$ and (b) $\Nuptotal$ as a function of the temperature for a
$10$~ns Gaussian 
pulse with peak intensities 
$\Ialign=\SI{e12}{\intensity}$  (thick solid line), $\Ialign=\SI{5e11}{\intensity}$ (dashed line) 
and $\Ialign=\SI{e11}{\intensity}$(doted line). The field  configuration is  
$\Estatabs=\SI{2}{k\fieldstrength}$ and $\beta=30\degree$. }
\end{figure}

In \autoref{fig:fc_2000_10_30_I},  we plot $\costT$ and $\Nuptotal$ 
for a thermal sample exposed to a $10$~ns pulse
combined with a dc field of $\Estatabs=\SI{2}{k\fieldstrength}$ tilted an angle $\beta=30\degree$. 
For cold samples with $T\lesssim 0.74$~K and $T\lesssim 0.69$~K, 
we obtain $\costT\gtrsim 0.5$ with $\Ialign=\SI{e12}{\intensity}$ and $\SI{5e11}{\intensity}$,
respectively.
For $\Ialign=\SI{5e11}{\intensity}$ and $\SI{e12}{\intensity}$,
we obtain $\Nuptotal\gtrsim 0.73$ if the 
rotational temperature is $T\lesssim1.1$~K. Thus, using this strong dc field
the  orientation  of a thermal ensemble becomes comparable to the experimental value for a 
quantum-state selected molecular beam in a very weak electric field.
For this strong electric field, the orientation of the quantum-state selected  beam
is  $\Nuptotal= 0.99$ using a  probe pulse linearly polarized along the vertical axis of the detector.

\begin{figure}[b]
\centering
\includegraphics[width=30pc]{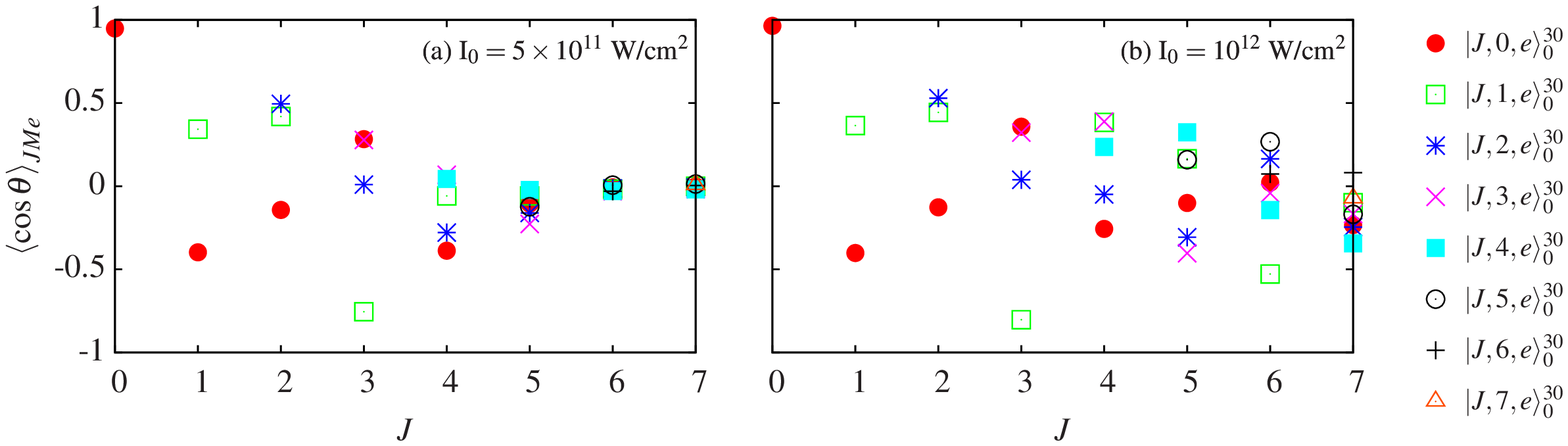}
\caption{\label{fig:cos_JM_2000_10_30} 
For the states 
\pstatetreinta{J}{M}{e}$_0$, we present the 
orientation cosines $\cost_{JMl}$ at $t=0$ versus the field-free rotational quantum number. 
The FWHM of the Gaussian pulses is $\tau=10$~ns  and the peak intensities
(a) $\Ialign=\SI{5e11}{\intensity}$ and
(b) $\Ialign=\SI{e12}{\intensity}$.
 The field  configuration is  $\Estatabs=\SI{2}{k\fieldstrength}$ and $\beta=30\degree$. }
\end{figure}

In \autoref{fig:cos_JM_2000_10_30}, we present the expectation value
$\cost_{JMl}$ at $t=0$ for several rotational states in $10$~ns Gaussian pulses
with $\Ialign=\SI{5e11}{\intensity}$ and  $\SI{e12}{\intensity}$,
$\Estatabs=\SI{2}{k\fieldstrength}$ and $\beta=30\degree$.
For both field configurations, the 
$\pstatetreinta{0}{0}{e}_0$ and $\pstatetreinta{3}{1}{e}_0$ states are strongly oriented and
antioriented, respectively. 
The remaining states show a moderate or weak orientation. 
The effect of doubling the peak intensity is not noticeable
for the levels with field-free rotational quantum number $J\le 3$, 
and, in addition, for a certain peak intensity, we encounter similar orientation using 
a Gaussian pulse of  $10$~ns or $5$~ns.
The rotational dynamics of the ground state is adiabatic for both pulses; whereas
for the excited state, this phenomenon can be explained by the non adiabatic effects taking place at
weak laser intensities. 
When the levels in a certain $J$ manifold are 
driven apart by the laser field, the process is nonadiabatic and 
there is a population transfer between them, already at weak laser intensities. 
Thus, the wave function of 
any excited level has contributions from adiabatic states
which correspond to different pendular doublets.
By further increasing the laser intensity, the molecular dynamics is affected
by the 
avoided crossings with adjacent levels having different field-free
magnetic quantum numbers $M$ and by the formation of these pendular doublets.
The rotational dynamics in most of these crossings will be nonadiabatic and has to be analyzed for
each specific state. 
When the electric field is strong, the energy splitting within the states in the pendular pair is sufficiently
large, and, as a consequence, the population transfer when the doublets are formed is not significant. 


For completeness,  in~\autoref{fig:coef_JM_300_2000}(b) 
we present the field-dressed rotational dynamics of the
state \pstatetreinta{2}{0}{e}$_t$ in a $10$~ns pulse with $\Ialign=\SI{e12}{\intensity}$
and a strong dc field of $\Estatabs=\SI{2}{k\fieldstrength}$. 
After an adiabatic  switching on of the electric field, the states in the $J=2$ manifold
are driven apart,  $|C_{20e}(t)|^2$ decreases as $\textup{I}(t)$ is increased,
whereas $|C_{21e}(t)|^2$  and $|C_{22e}(t)|^2$ increase.
Compared to the weak dc field case in~\autoref{fig:coef_JM_300_2000}(a), this $J$-manifold 
splitting 
takes place at a stronger laser intensity, because the energy gap between the adiabatic states 
\ppstatetreinta{2}{2}{e}, \ppstatetreinta{2}{1}{e} and \ppstatetreinta{2}{0}{e} 
is larger for $\Estatabs=\SI{2}{k\fieldstrength}$ than for $\Estatabs=\SI{300}{\fieldstrength}$. 
Let us mention that by further increasing  $\Estatabs$, the energy splitting within this 
$J$-manifold is increased, and, therefore, this population redistribution will be 
reduced~\cite{omiste_nonadiabatic_2012}.
The avoided crossing between the states \ppstatetreinta{2}{1}{e} and
\ppstatetreinta{2}{2}{e} occurs at $\textup{I}(t)\approx\SI{2.96e10}{\intensity}$, whereas the
one involving the levels \ppstatetreinta{2}{0}{e} and \ppstatetreinta{3}{3}{e} 
around  $\textup{I}(t)\approx\SI{1.09e11}{\intensity}$. Again, both of them are crossed diabatically, and the
population of the adiabatic states is interchanged.
By further increasing $\textup{I}(t)$, the pendular doublets start to form. In this case, 
the dc field is stronger and 
the energy gap is larger but the coupling due to the ac field is the same, then 
the population transfer is reduced. Indeed,
the adiabatic states \ppstatetreinta{1}{0}{e}, \ppstatetreinta{2}{0}{e} and
\ppstatetreinta{3}{2}{e}, the partners in the pendular doublets of 
\ppstatetreinta{2}{2}{e}, \ppstatetreinta{2}{1}{e} and
\ppstatetreinta{3}{3}{e}, respectively, show a small population, which is below $0.01$ 
once the peak intensity at $t=0$ is achieved. 
Thus, the population  at $t=0$ for the field-dressed state $\pstatetreinta{2}{0}{e}_0$  is 
$|C_{22e}(0)|^2=0.56$, 
$|C_{21e}(0)|^2=0.39$, 
and 
$|C_{33e}(0)|^2=0.05$.
These results are similar for the four pulses formed by combining 
$\tau =5$~ns and $10$~ns with $\Ialign=\SI{5e11}{\intensity}$ and  $\SI{e12}{\intensity}$.

\section{Conclusions}
\label{sec:conclusions}
In this work, we investigate the mixed-field orientation dynamics of a thermal sample of linear molecules.
We solve the time-dependent Schr\"odinger equation within the rigid rotor approximation for a
large set of rotational states. 
 As prototype example, we use the  OCS molecule. 
However,
 we stress that the above results could be used to describe the mixed-field orientation of 
  a thermal ensemble  of other polar linear molecules  by rescaling the Hamiltonian
  \eqref{eq:hamiltonian} in terms of the rotational constant.
 
By considering prototypical field configurations with weak dc fields, as in current mixed-field
orientation experiments, we have proven that the rotational temperature of the molecular beam should
be smaller than $0.7$~K to achieve a significant orientation.
Using a weak electric field, if the aim is a 
strongly oriented molecular ensemble, this should be as pure as possible in the ground state.
Thus, it is required a  quantum-state-selected
molecular beam,  unless the rotational temperature could be
efficiently reduced below $1$~K. 
It is found that a significant orientation  is
achieved for $1$~K molecular samples when 
the electric field strength is increased. 


\begin{acknowledgments}
Financial support by the Spanish project FIS2011-24540 (MICINN), the Grants
 P11-FQM-7276 and FQM-4643 (Junta de Andaluc\'{\i}a), and the
 Andalusian research group FQM-207 is
gratefully appreciated. J.J.O. acknowledges the support of ME under the program FPU.
\end{acknowledgments}

\section*{References}

\bibliographystyle{iopart-num}


\begin{thebibliography}{10}
\expandafter\ifx\csname url\endcsname\relax
  \def\url#1{{\tt #1}}\fi
\expandafter\ifx\csname urlprefix\endcsname\relax\def\urlprefix{URL }\fi
\providecommand{\eprint}[2][]{\url{#2}}

\bibitem{friedrich:jcp111}
Friedrich B and Herschbach D~R 1999 {\em J. Chem. Phys.\/} {\bf 111} 6157

\bibitem{friedrich:jpca103}
Friedrich B and Herschbach D 1999 {\em J. Phys. Chem. A\/} {\bf 103} 10280

\bibitem{nielsen:prl2012}
Nielsen J~H, Stapelfeldt H, K\"upper J, Friedrich B, Omiste J~J and
  Gonz\'alez-F\'erez R 2012 {\em Phys. Rev. Lett.\/} {\bf 108} 193001

\bibitem{omiste_nonadiabatic_2012}
Omiste J~J and Gonz\'alez-F\'erez R 2012 {\em Phys. Rev. A\/} {\bf 86} 043437

\bibitem{sakai:prl_90}
Sakai H, Minemoto S, Nanjo H, Tanji H and Suzuki T 2003 {\em Phys. Rev.
  Lett.\/} {\bf 90} 083001

\bibitem{Buck:IRPC25:583}
Buck U and F\'arn\'ik M 2006 {\em Int.\ Rev.\ Phys.\ Chem.\/} {\bf 25} 583

\bibitem{ghafur_impulsive_2009}
Ghafur O, Rouzee A, Gijsbertsen A, Siu W~K, Stolte S and Vrakking M~J~J 2009
  {\em Nat Phys\/} {\bf 5} 289--293

\bibitem{kupper:prl102}
Holmegaard L, Nielsen J~H, Nevo I, Stapelfeldt H, Filsinger F, K\"upper J and
  Meijer G 2009 {\em Phys. Rev. Lett.\/} {\bf 102} 023001

\bibitem{kupper:jcp131}
Filsinger F, K\"upper J, Meijer G, Holmegaard L, Nielsen J~H, Nevo I, Hansen
  J~L and Stapelfeldt H 2009 {\em J. Chem. Phys.\/} {\bf 131} 064309

\bibitem{even:8068}
Even U, Jortner J, Noy D, Lavie N and Cossart-Magos C 2000 {\em J. Chem.
  Phys.\/} {\bf 112} 8068

\bibitem{seideman2006}
Seideman T and Hamilton E 2006 {\em Adv. Atom. Mol. Opt. Phys.\/} {\bf 52} 289

\bibitem{feit:jcp82}
Feit M~D, {Fleck Jr} J~A and Steiger A 1982 {\em J. Comp. Phys.\/} {\bf 47} 412

\bibitem{bacic:arpc89}
Ba{\u{c}i\`c} Z and Light J~C 1989 {\em Annu. Rev. Phys. Chem.\/} {\bf 40} 469

\bibitem{corey:jcp92}
Corey G~C and Lemoine D 1992 {\em J. Chem. Phys.\/} {\bf 97} 4115

\bibitem{offer:10416}
Offer A~R and Balint-Kurti G~G 1994 {\em J. Chem. Phys.\/} {\bf 101}
  10416--10428

\bibitem{sanchezmoreno:PRA.2007}
S\'anchez-Moreno P, Gonz\'alez-F\'erez R and Schmelcher P 2007 {\em Phys. Rev.
  A\/} {\bf 76} 053413

\bibitem{hartelt_jcp128}
H\"artelt M and Friedrich B 2008 {\em J. Chem. Phys.\/} {\bf 128} 224313

\bibitem{omiste:pccp2011}
Omiste J~J, G\"arttner M, Schmelcher P, Gonz\'{a}lez-F\'{e}rez R, Holmegaard L,
  Nielsen J~H, Stapelfeldt H and K\"upper J 2011 {\em Phys. Chem. Chem.
  Phys.\/} {\bf 13} 18815--18824

\end{thebibliography}

\end{document}